\documentclass{aa}

\usepackage{graphicx,natbib}
\usepackage{bm}
\usepackage{amsmath}

\usepackage{color}
\usepackage{ulem}

\begin{document}
   \title{Hot-cold plasma transition region: collisionless case}

   \authorrunning{Karlick\'y and Karlick\'y}
   \titlerunning{Hot-cold plasma transition region}

   \author{M. Karlick\'y$^{1}$ and F. Karlick\'y$^2$
          }
   \offprints{, \email{karlicky@asu.cas.cz}}

   \institute{Astronomical Institute of the Czech Academy of Sciences, Fri\v{c}ova 298, CZ -- 251 65 Ond\v rejov, Czech Republic
         \and
   Department of Physics, Faculty of Science, University of Ostrava, 30~dubna 22, 701 03 Ostrava, Czech Republic\\}
   \date{Received ; accepted }


  \abstract
   {}
   {We study processes at the transition region between hot (rare) and cold (dense) plasma
   in the collisionless regime.}
   {We use a 3-dimensional electromagnetic particle-in-cell (3-D PIC) relativistic code.}
   {We initiate a transition region between the hot (rare) and cold (dense) plasma.
   Motivated by the transition region
   in the solar atmosphere the temperature and density ratio of the plasmas is chosen
   as 100 and 0.01, respectively. For better understanding of studied processes
   we make two types of computations: a) without any
   interactions among plasma particles (free expansion) and b) with the full electromagnetic interactions (but
   no particle-particle collisions). In both the cases we found that the flux of cold plasma electrons and protons
   from colder plasma to hotter one dominates over the flux of hot plasma electrons and protons in the opposite direction.
   Thus, the plasma in the hotter part of the system becomes colder and denser during time evolution.
   In the case without any interactions
   among particles the cold plasma electrons and protons freely penetrate into the hot plasma. But,
   the cold plasma electrons are faster than cold plasma protons and therefore they penetrate deeper into the hotter part of the system
   than the protons.  Thus, the cooling of the electron and proton components of the plasma in the hotter part of the system
   is different. On the other hand, in the case with the electromagnetic interactions, owing to the plasma property,
   which tries to keep the total electric current constant everywhere (close to zero in our case),
   the cold plasma electrons penetrate into the hotter part of the system together with the cold plasma protons. Thus, the cooling
   of the hotter plasma is much slower and it is the same for electrons and protons.  Moreover, the plasma waves generated
   at the transition region during
   these processes reduce the number of electrons escaping from the hot plasma into the
   colder one and scatter the cold plasma particles penetrating to the hot plasma. Therefore these waves
   support a temperature jump between hot and cold plasma. Nevertheless, for
   stabilizing of this temperature jump as in the real solar transition region, we propose that the
   gravity force and more or less permanent heating source in the corona
   (nanoflares?) together with the appropriate radiative losses at both sides of the transition region need to be included.
   In our case it is interesting that just in front of the expanding cold plasma to hotter one
   the formed electric field is opposite to that found in the
   case of the plasma expansion into vacuum or in the case of the plasma expansion to plasma with lower density,
   but having the same temperature. We propose that it is caused by a high temperature difference between
   hot and cold plasmas at the transition region.}
   {}


   \maketitle

\section{INTRODUCTION}

In the solar atmosphere there is a layer between the relatively cool
chromosphere ($\approx$ 10000 K) and hot corona ($\ge$ 1 MK) which is called
the transition region. Although in models of the solar atmosphere the
transition region is described as the smooth interface between the chromosphere
and corona, in reality it is a highly dynamic place, where a cool chromospheric
material protrudes (such as prominences) or shoots out into the corona (as
observed in spicules and surges)
\citep{1992str..book.....M,2002AdSpR..30..501P,2002ESASP.508..237P,2008ApJ...683L..87J,2009MmSAI..80..654Z}.

The problem of the solar transition region is closely connected with the
problem of the coronal heating. Many models explaining the coronal heating were
proposed: a) stressing and reconnection models
\citep[e.g.][]{1988ApJ...330..474P,1992ApJ...390..297H,1999ApJ...521..451S,2002ApJ...572L.113G},
wave models
\citep[e.g.][]{1981ARA&A..19....7K,1988JGR....93.9547H,1995SoPh..157...75G,2002ApJ...575..571D},
and velocity filtration models \citep[e.g.][]{1994ApJ...427..446S}. Moreover,
\cite{2015ApJ...810L...1S} presented the model, where the electron accelerated
in the chromospheric magnetic reconnection are further accelerated in double
layers (formed in an expanding plasma) and thus heat the corona.

There are also many magnetohydrodynamic models of the transition region:
static, steady flow or dynamic models, for their review, see the book of
\cite{1992str..book.....M}. Such types of models were further evolved. For
example, \cite{2011A&A...531A..97Z} compared model velocities with the observed
Doppler shifts of the transition emission lines. Furthermore,
\cite{2012AstL...38..801P} tried to explain the observed extreme ultraviolet
radiation from the transition region, and
\cite{2015CEAB...39...43B,2015AstL...41..601B} studied effects of the anomalous
thermal conductivity and collective plasma processes on the structure of the
transition region.

However, the transition region is very narrow. Its thickness is estimated as a
few hundred kilometers, which is comparable to Coulomb mean free path for
electrons and protons \citep{1993ASSL..184.....B}. Therefore, collisionless
effects can be expected.

Due to the narrowness and highly dynamic state of the transition region the
processes forming the transition region are not still well understood
\citep{2014nest.book.....G}. Moreover, considering the problem of thermal
fronts in astrophysical plasmas \citep{2015ApJ...814..153K}, there appears a
question if the solar transition region is not some kind of the thermal front
or cascade of thermal fronts.

For these reasons, in the present paper we study processes in in the hot-cold
transition region using the 3-dimensional electromagnetic particle-in-cell
(PIC) relativistic model, where only collisionless processes are included. We
hope that such a study helps to understand processes in the transition region.

The paper is structured as follows: In Section 2 the numerical model is described. The results are presented in section 3,
Finally, discussion and conclusions are in Section 4.

\begin{figure}
\begin{center}
\includegraphics[width=8cm]{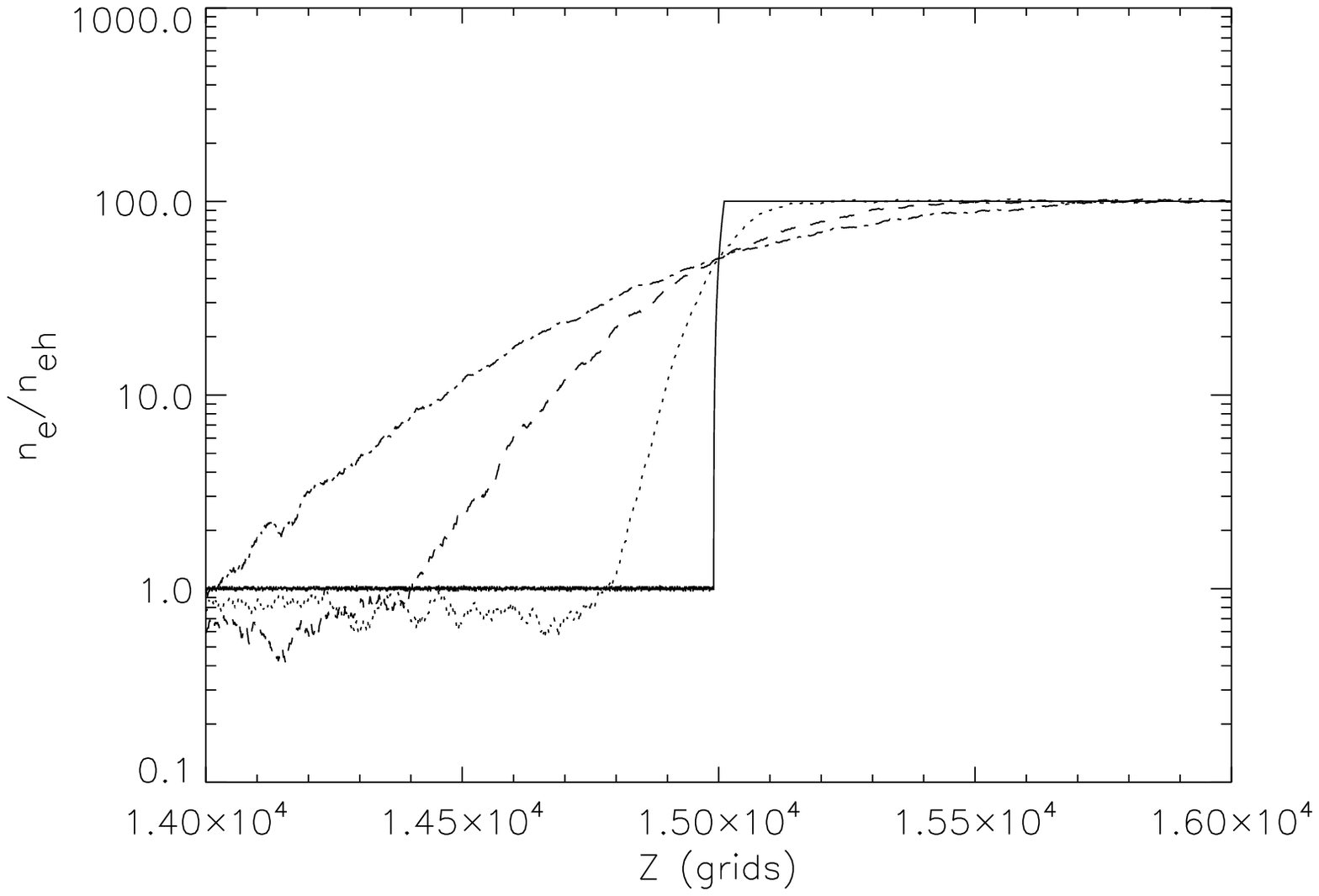}
\includegraphics[width=8cm]{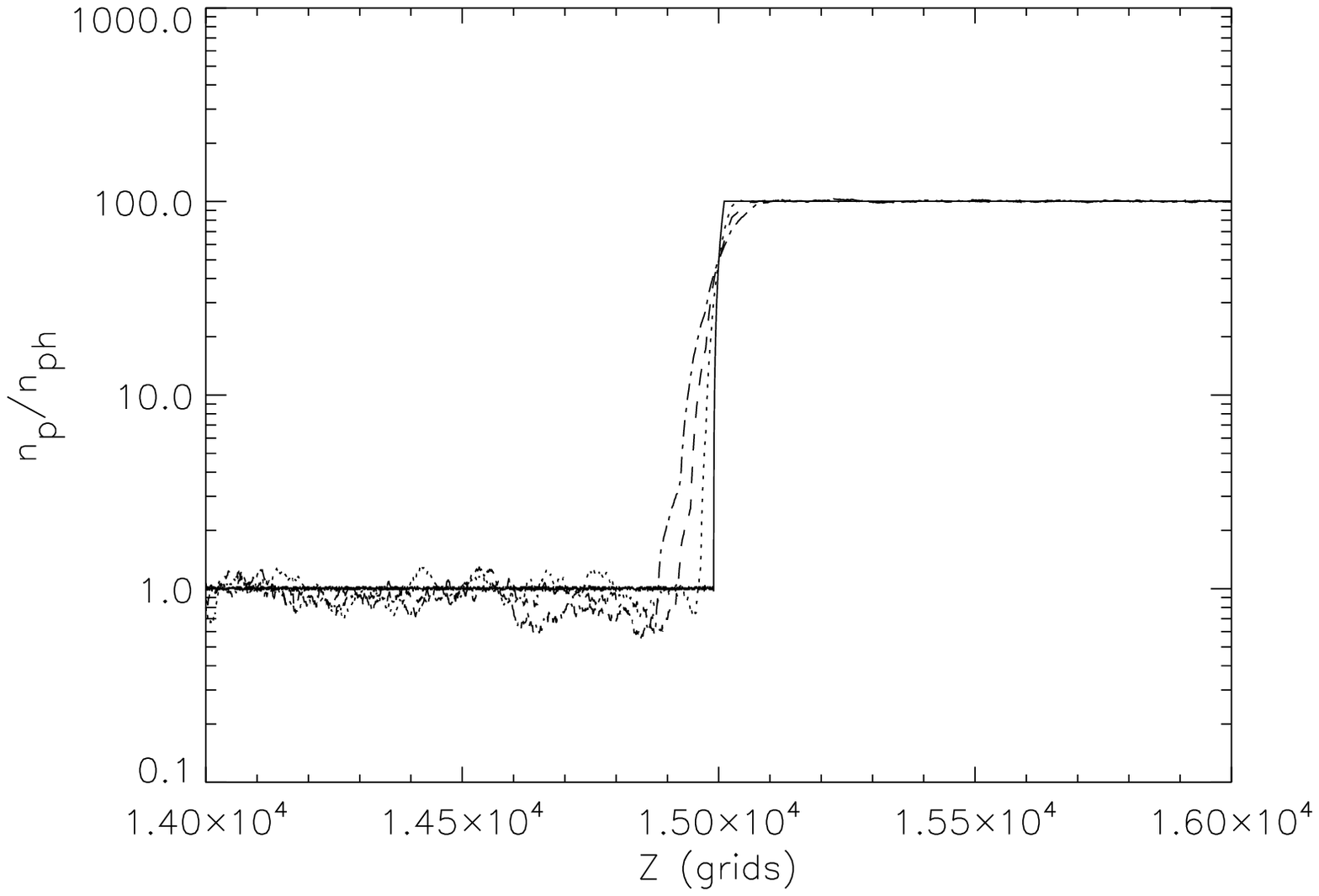}
\end{center}
  \caption{Upper part: Ratio of the electron density $n_e$ to that of the hot plasma $n_{eh}$
           along the z-coordinate at the initial state (solid line),
   at $\omega_{pe}t$ = 500 (dotted line), at $\omega_{pe}t$ = 1500 (dashed line),  and at $\omega_{pe}t$ = 2500
   (dash-dot line).
   Bottom part: The same for protons. (Run with the free expansion of particles.)}
  \label{figure1}
\end{figure}

\begin{figure}
\begin{center}
\includegraphics[width=8cm]{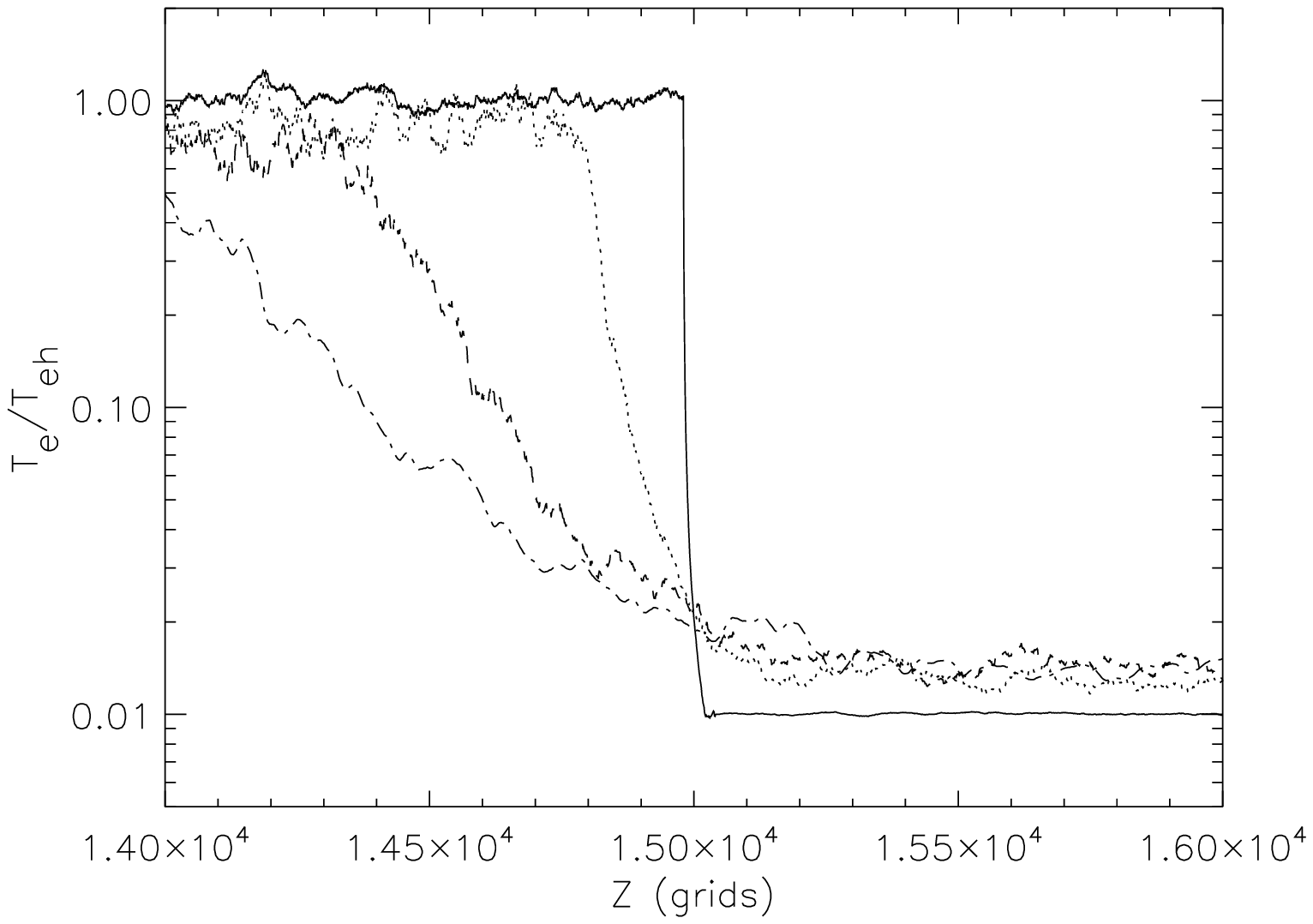}
\includegraphics[width=8cm]{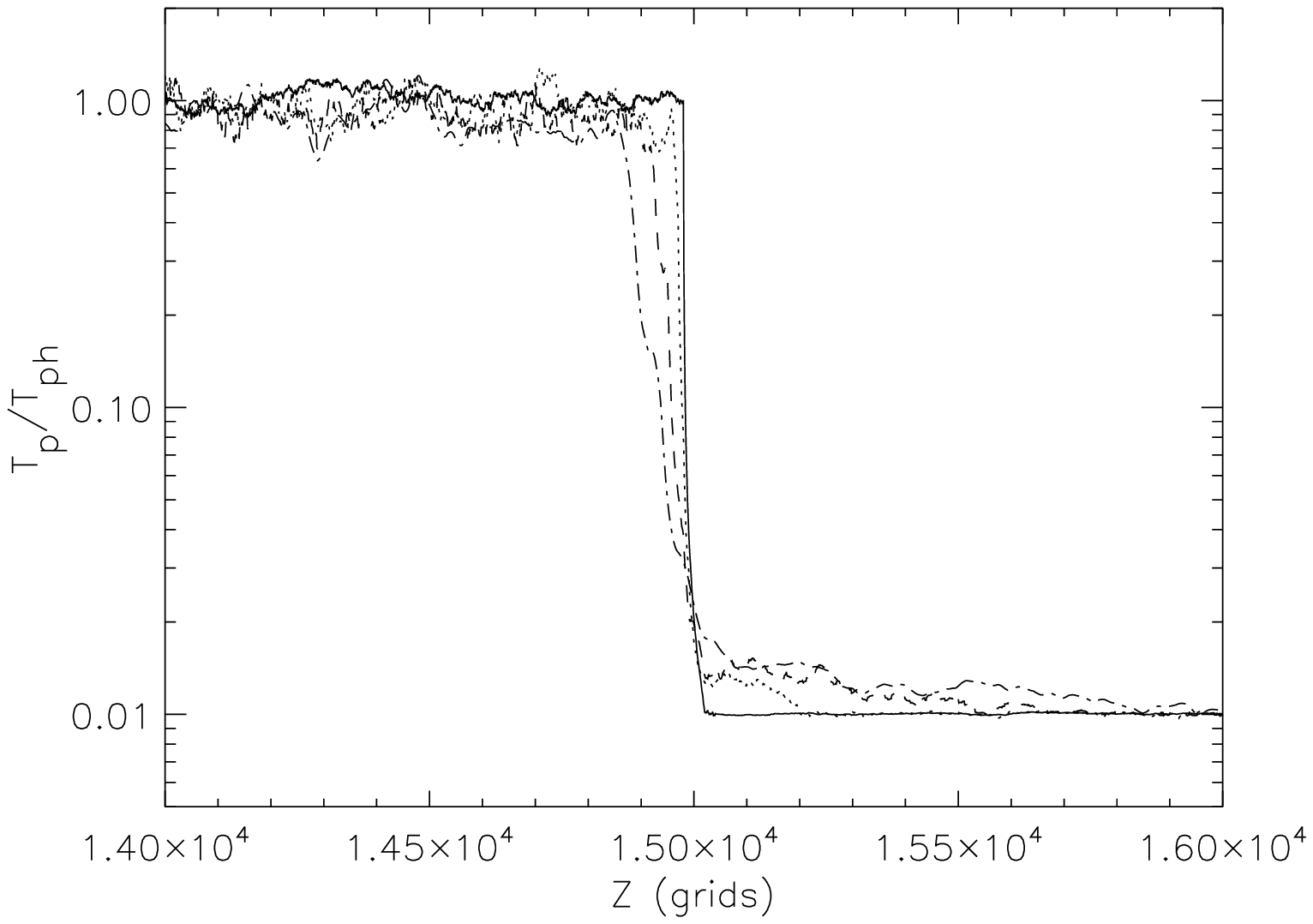}
\end{center}
  \caption{Upper part: Ratio of the electron kinetic energy $T$ (pseudo-temperature) to that of the hot plasma $T_h$
along the z-coordinate at the initial state (solid line),
   at $\omega_{pe}t$ = 500 (dotted line), at $\omega_{pe}t$ = 1500 (dashed line),  and at $\omega_{pe}t$ = 2500
   (dash-dot line).
   Bottom part: The same for protons. (Run with the free expansion of particles.)}
  \label{figure2}
\end{figure}

\begin{figure}
\begin{center}
\includegraphics[width=8cm]{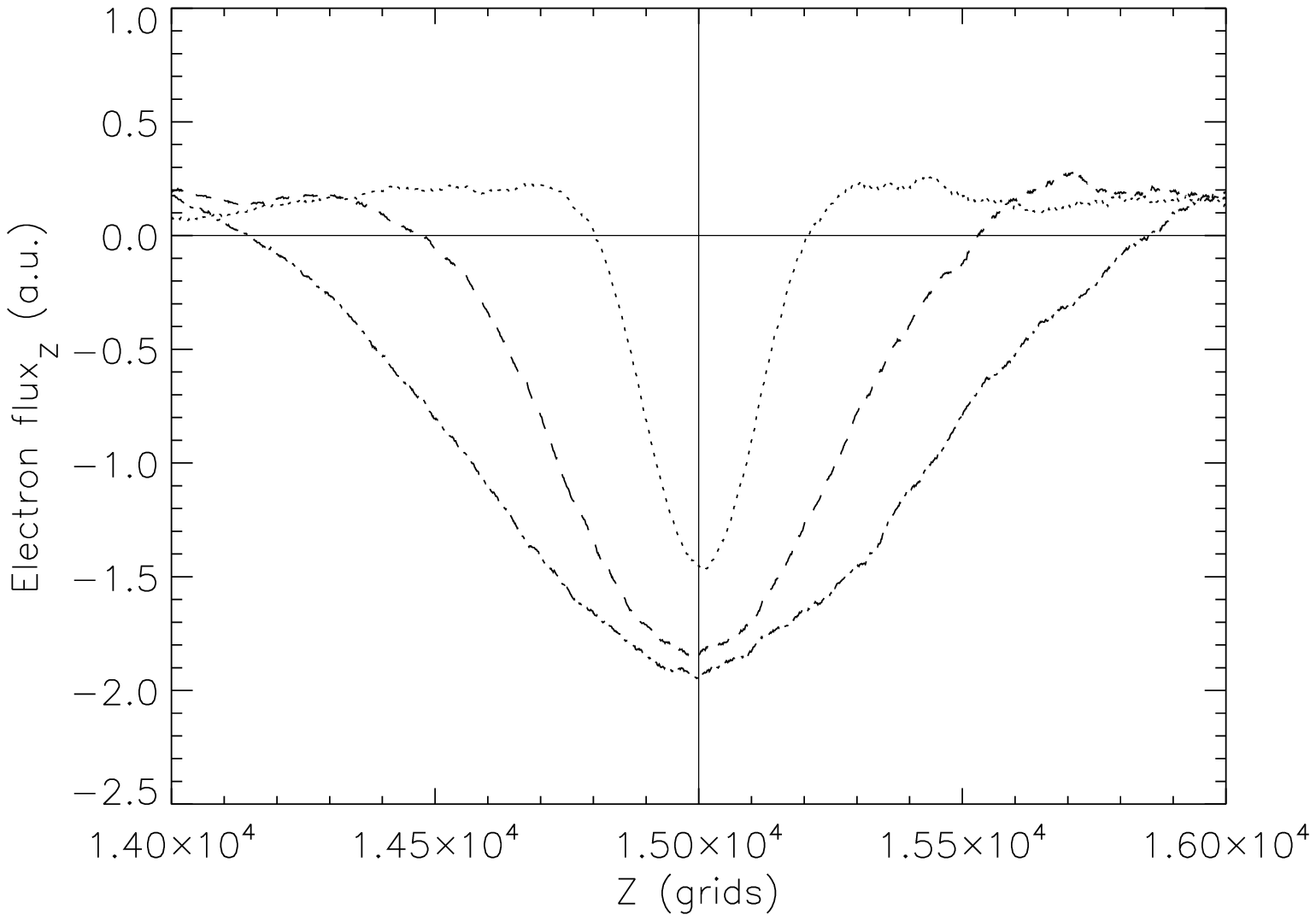}
\includegraphics[width=8cm]{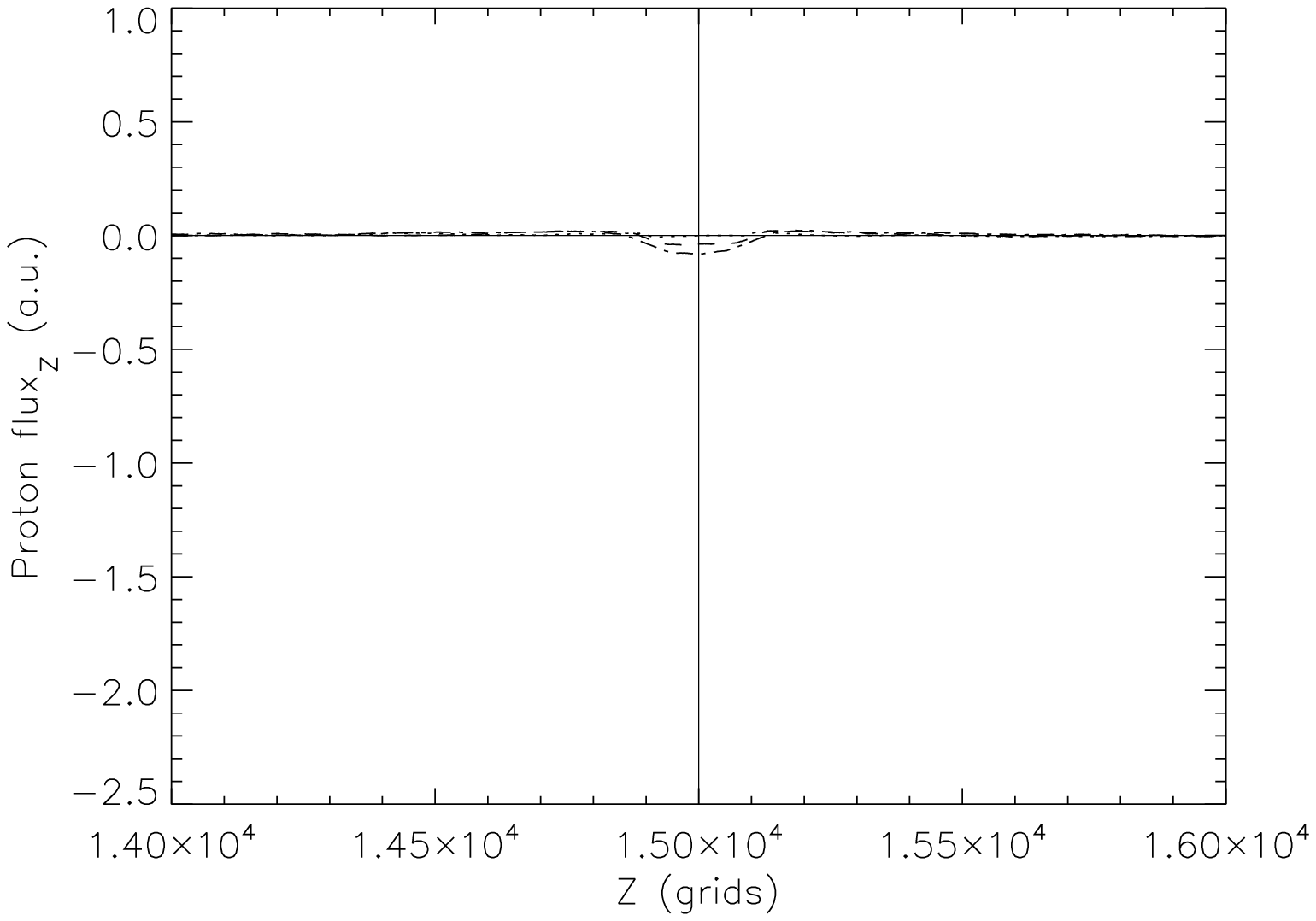}
\end{center}
  \caption{Upper part: Electron flux along the z-coordinate at $\omega_{pe}t$ = 500 (dotted line),
   at $\omega_{pe}t$ = 1500 (dashed line),  and at $\omega_{pe}t$ = 2500 (dash-dot line).
   Bottom part: The same for proton flux. (Run with the free expansion of particles.)}
  \label{figure3}
\end{figure}

\begin{figure}
\begin{center}
\includegraphics[width=8cm]{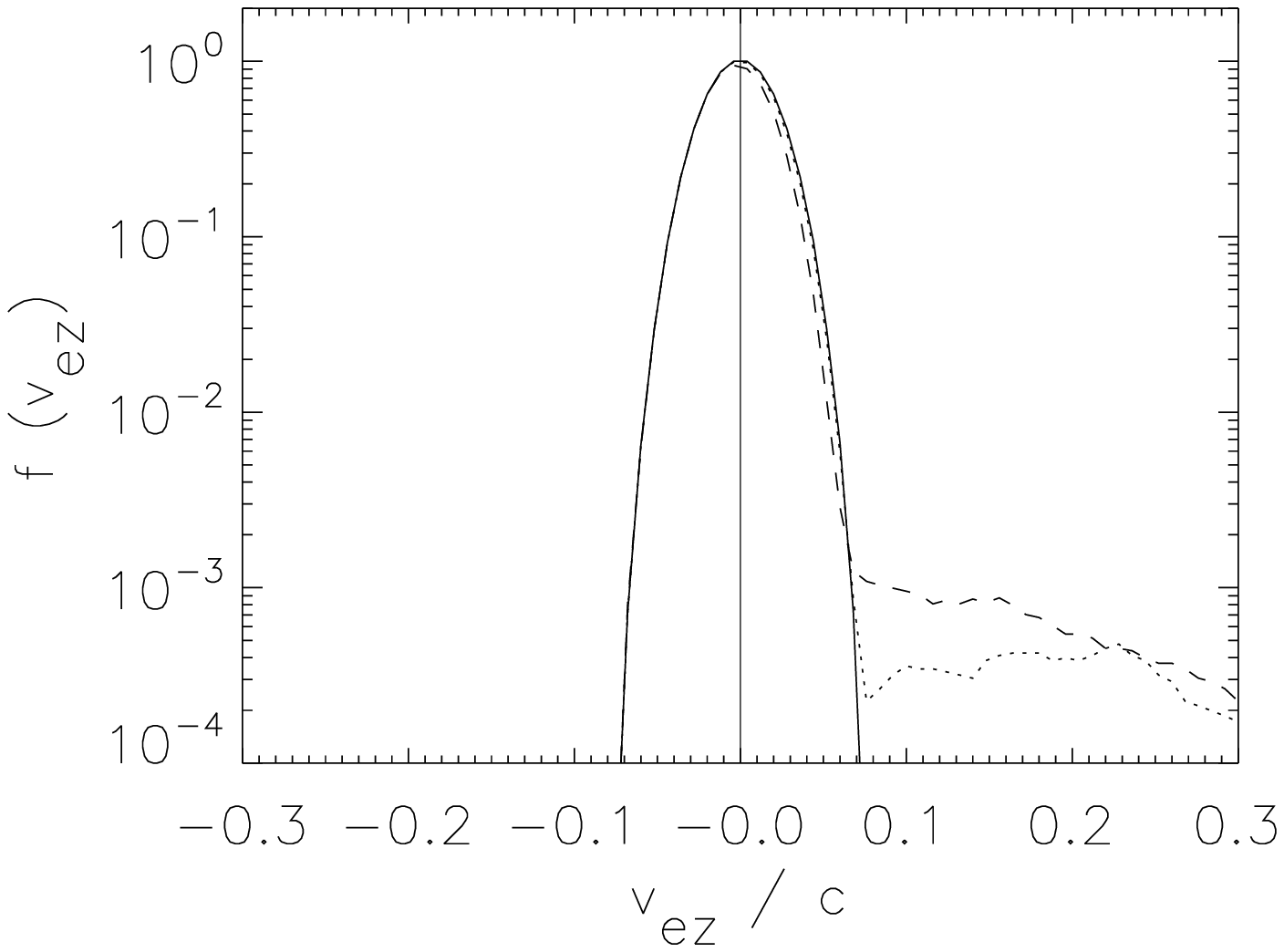}
\includegraphics[width=8cm]{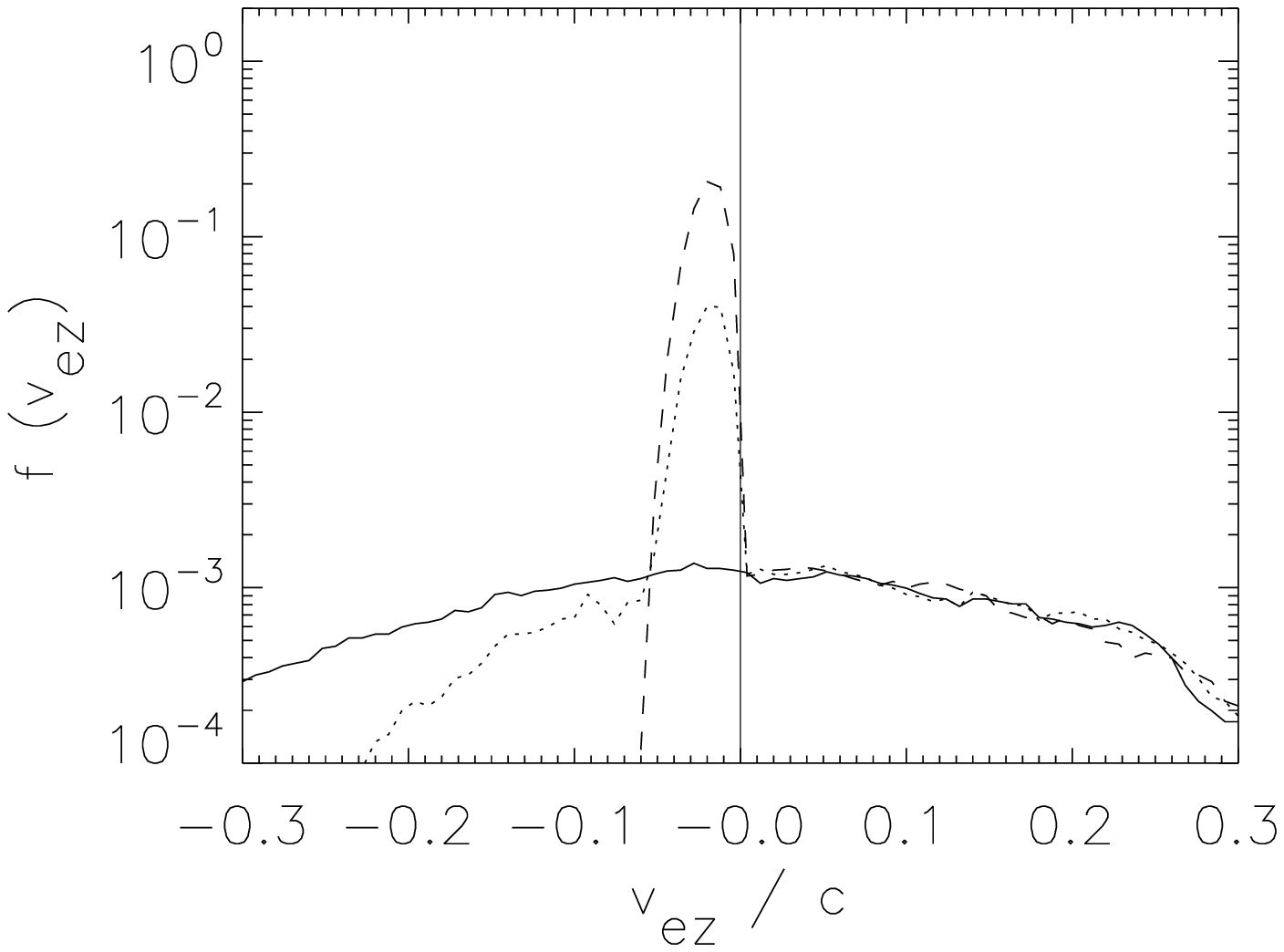}

\end{center}
  \caption{Upper part: Electron distribution  in the $v_{ez}$ velocity (normalized to the distribution maximum) in the z-direction (c means the speed of light)
  in the space interval 15000 - 16000 grids at the initial state (solid line),
   at $\omega_{pe}t$ = 500 (dotted line), and at $\omega_{pe}t$ = 2500 (dashed line).
   Bottom part: Electron distribution in the $v_{ez}$ velocity (normalized to the distribution maximum from the upper part) in the z-direction (c means the speed of light)
  in the space interval 14000 - 15000 grids at the initial state (solid line),
   at $\omega_{pe}t$ = 500 (dotted line), and at $\omega_{pe}t$ = 2500 (dashed line).
   (Run with the free expansion of particles.)}
  \label{figure4}
\end{figure}

\begin{figure}
\begin{center}
\includegraphics[width=8cm]{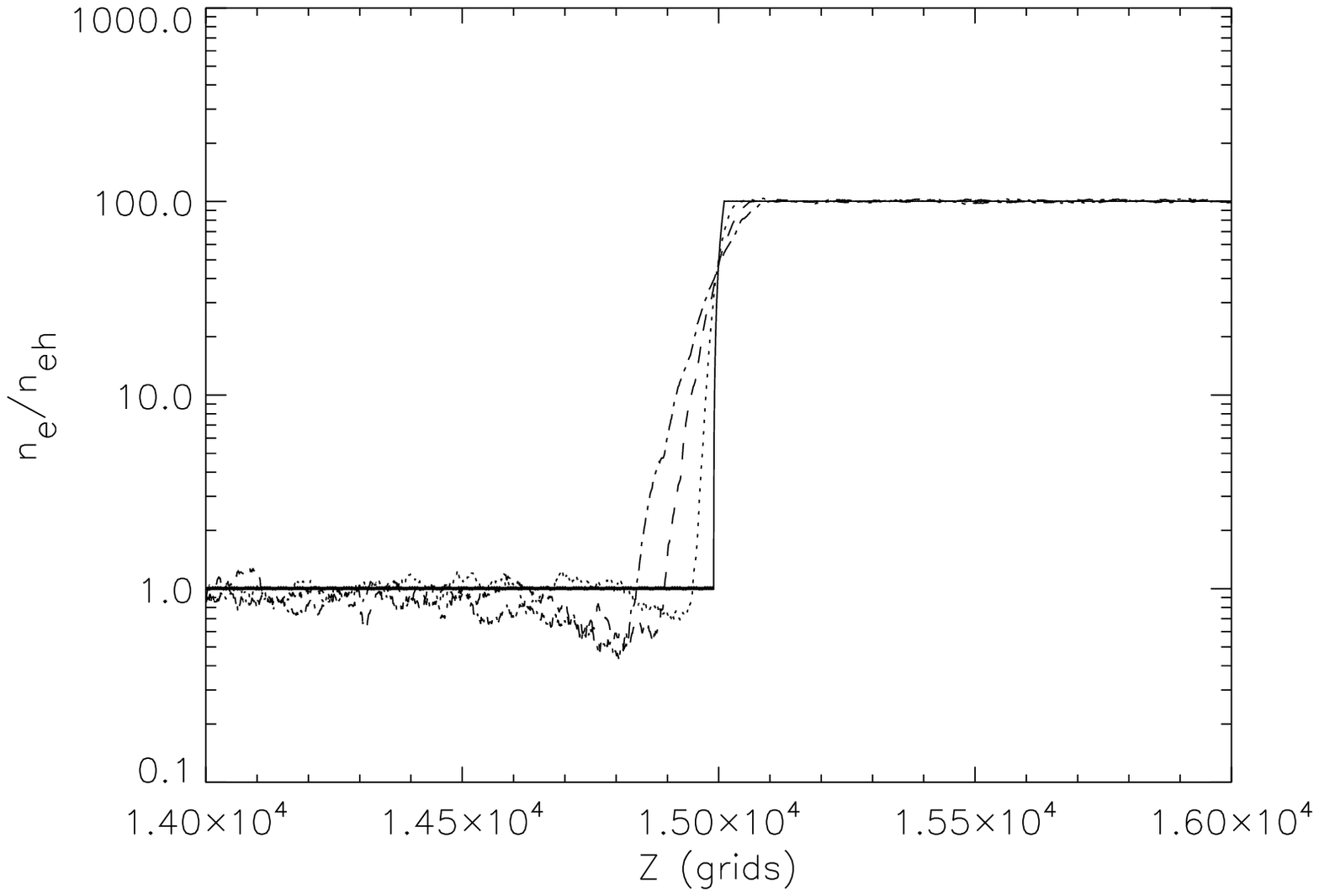}
\includegraphics[width=8cm]{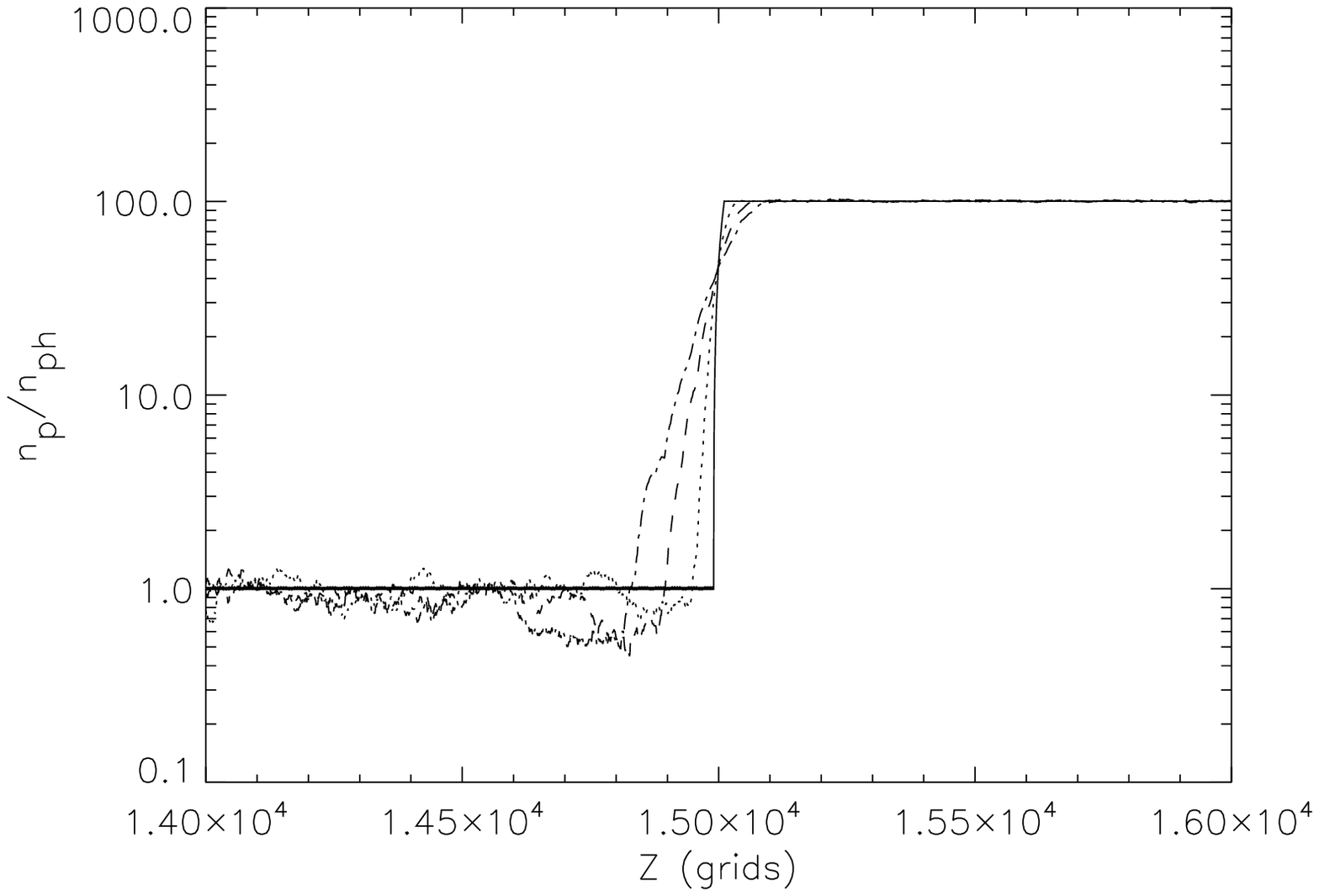}
\end{center}
  \caption{Upper part: Ratio of the electron density $n_e$ to that of the hot plasma $n_{eh}$
           along the z-coordinate at the initial state (solid line),
   at $\omega_{pe}t$ = 500 (dotted line), at $\omega_{pe}t$ = 1500 (dashed line),  and at $\omega_{pe}t$ = 2500
   (dash-dot line).
   Bottom part: The same for protons. (Run with electromagnetic interactions.)}
  \label{figure5}
\end{figure}

\begin{figure}
\begin{center}
\includegraphics[width=8cm]{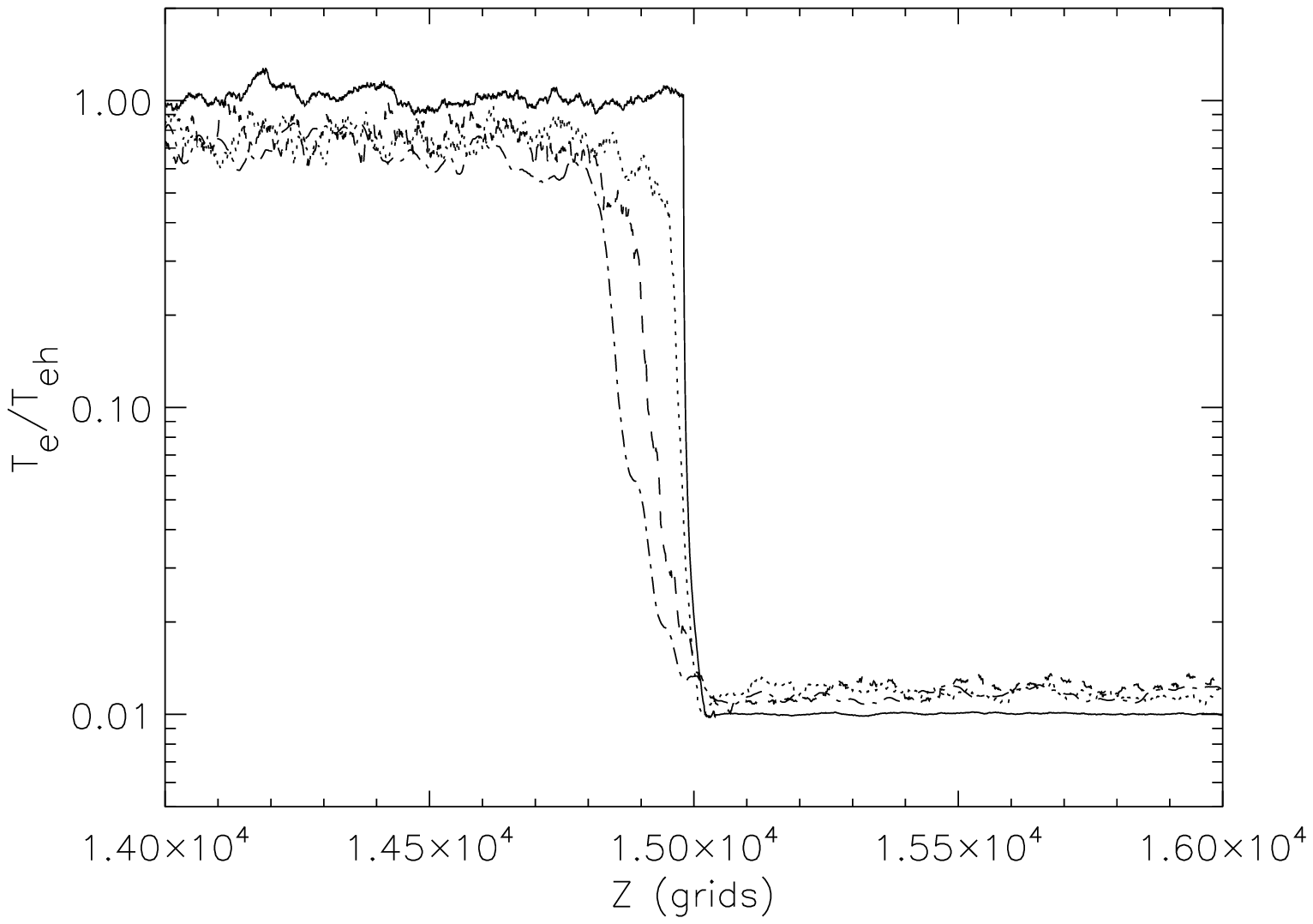}
\includegraphics[width=8cm]{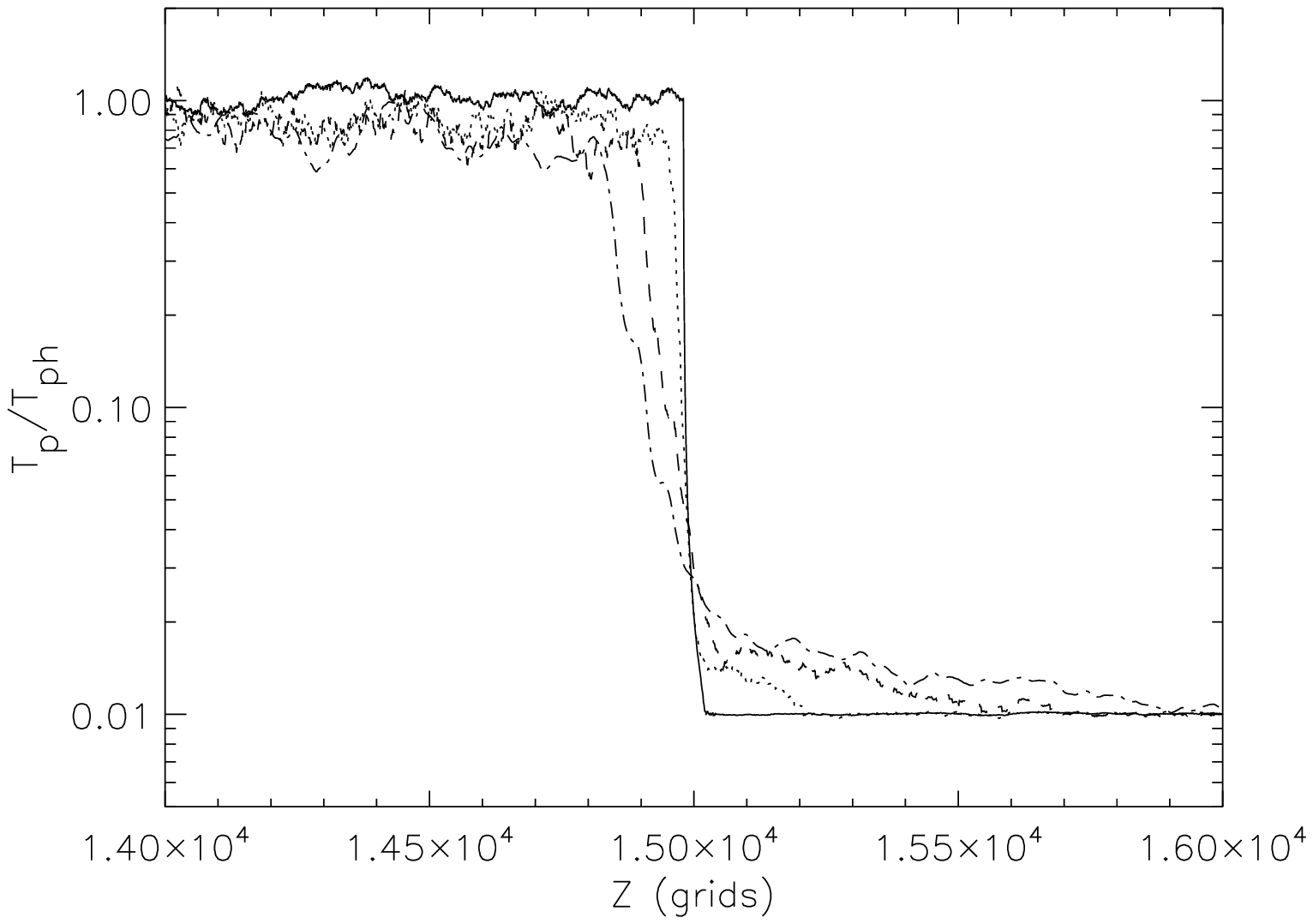}
\end{center}
  \caption{Upper part: Ratio of the electron kinetic energy $T$ (pseudo-temperature) to that of the hot plasma $T_h$
along the z-coordinate at the initial state (solid line),
   at $\omega_{pe}t$ = 500 (dotted line), at $\omega_{pe}t$ = 1500 (dashed line),  and at $\omega_{pe}t$ = 2500
   (dash-dot line).
   Bottom part: The same for protons. (Run with electromagnetic interactions.)}
  \label{figure6}
\end{figure}

\begin{figure}
\begin{center}
\includegraphics[width=8cm]{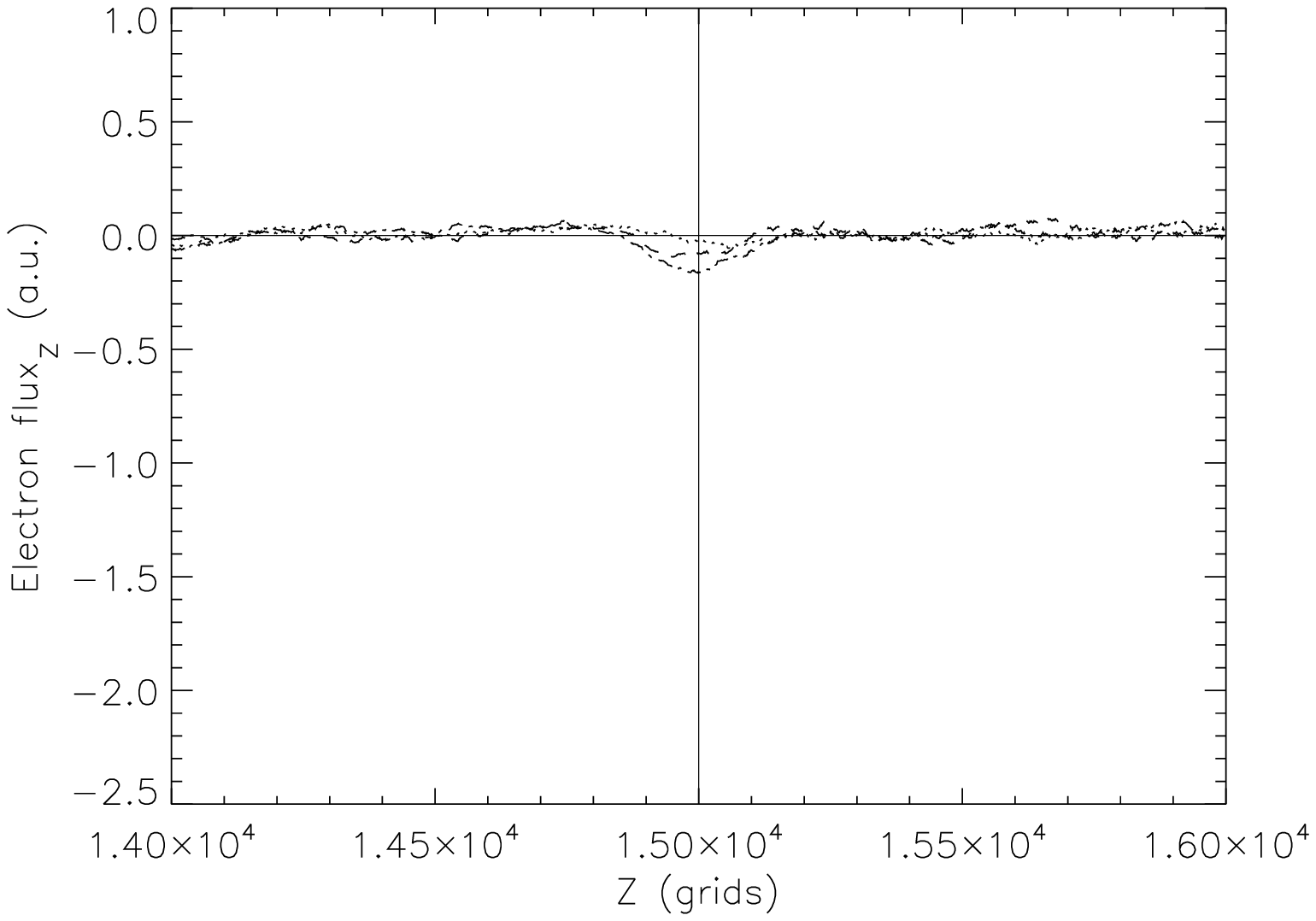}
\includegraphics[width=8cm]{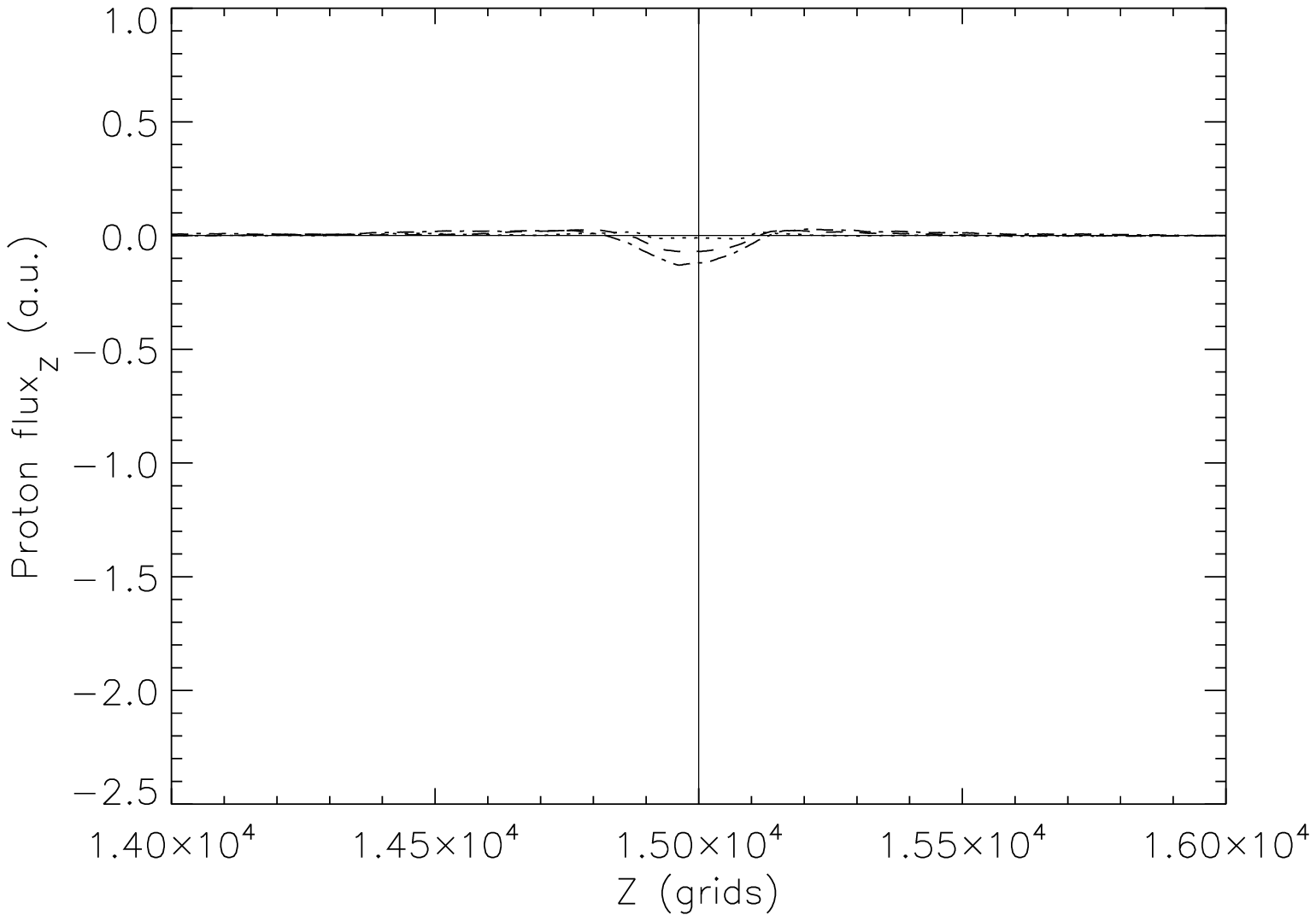}
\end{center}
  \caption{Upper part: Electron flux along the z-coordinate at $\omega_{pe}t$ = 500 (dotted line),
   at $\omega_{pe}t$ = 1500 (dashed line),  and at $\omega_{pe}t$ = 2500 (dash-dot line).
   Bottom part: The same for proton flux. (Run with electromagnetic interactions.)}
  \label{figure7}
\end{figure}

\begin{figure}
\begin{center}
\includegraphics[width=8cm]{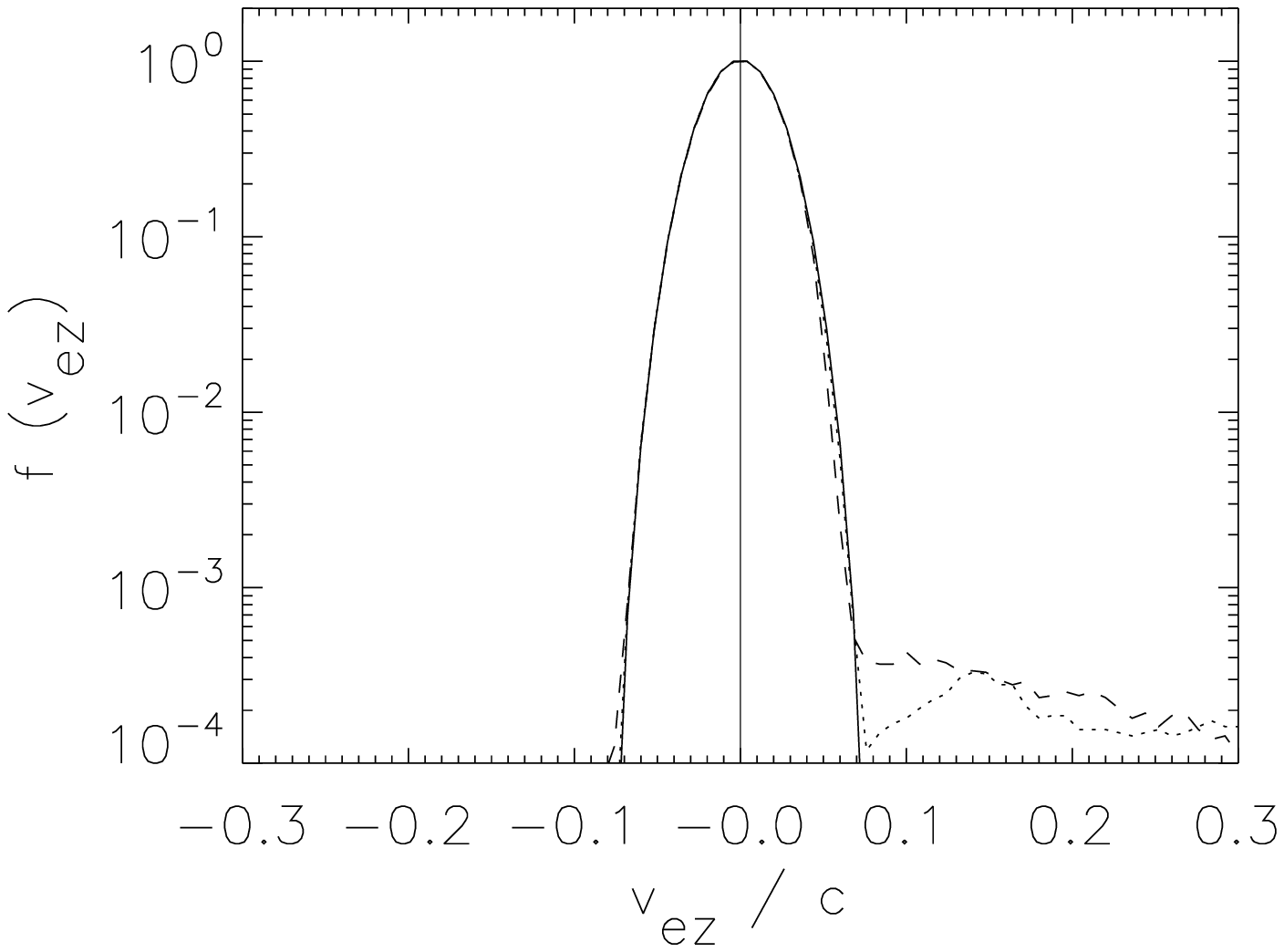}
\includegraphics[width=8cm]{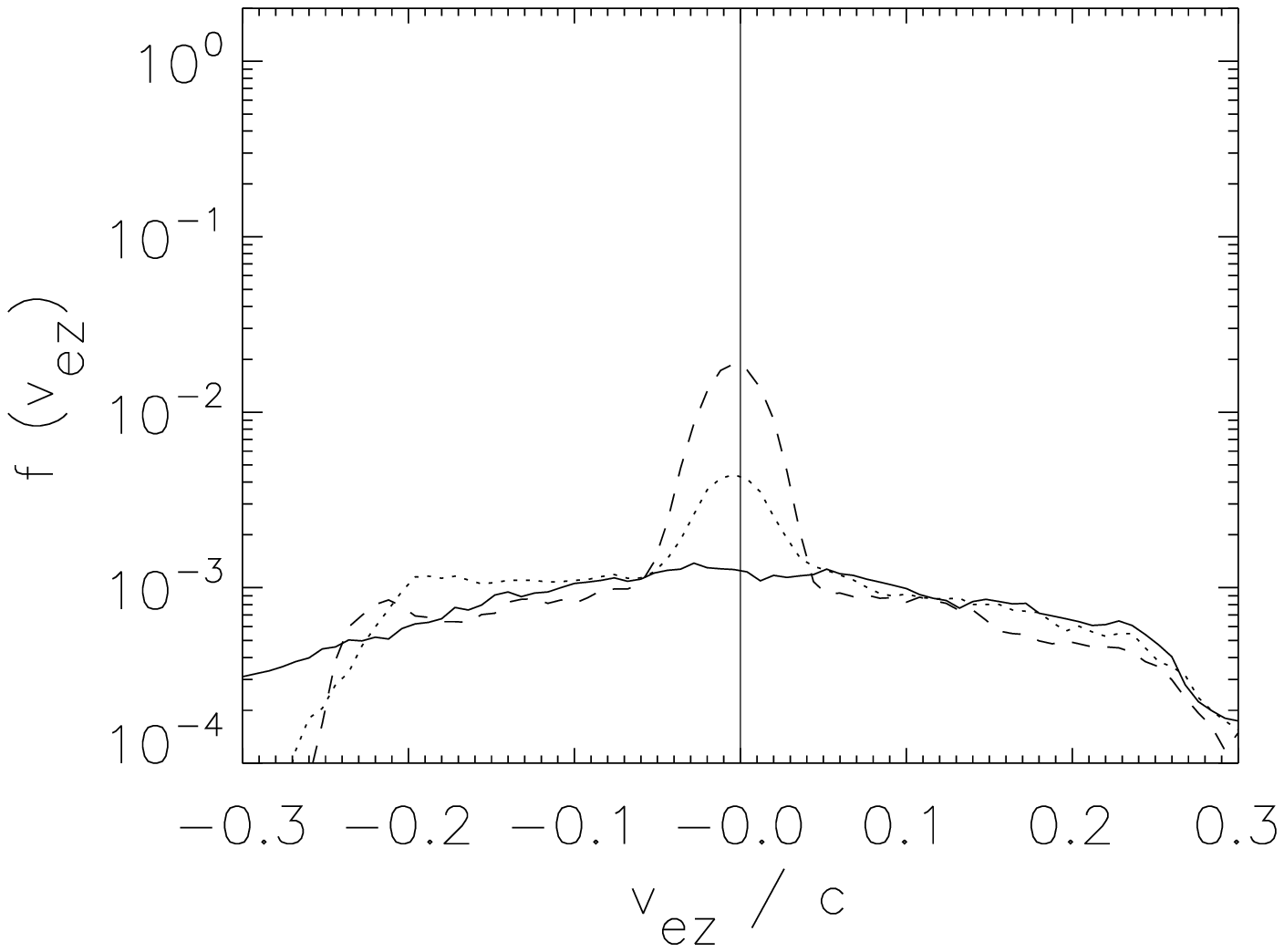}
\end{center}
  \caption{Upper part: Electron distribution in the $v_{ez}$ velocity (normalized to the distribution maximum, c means the speed of light)
  in the space interval 15000 - 16000 grids at the initial state (solid line),
   at $\omega_{pe}t$ = 500 (dotted line), and at $\omega_{pe}t$ = 2500 (dashed line).
   Bottom part: Electron distribution in $v_{ez}$ velocity (normalized to the distribution maximum from the upper part)
  in the space interval 14000 - 15000 grids at the initial state (solid line),
   at $\omega_{pe}t$ = 500 (dotted line), and at $\omega_{pe}t$ = 2500 (dashed line).
   (Run with electromagnetic interactions.)}
  \label{figure8}
\end{figure}

\begin{figure}
\begin{center}
\includegraphics[width=8cm]{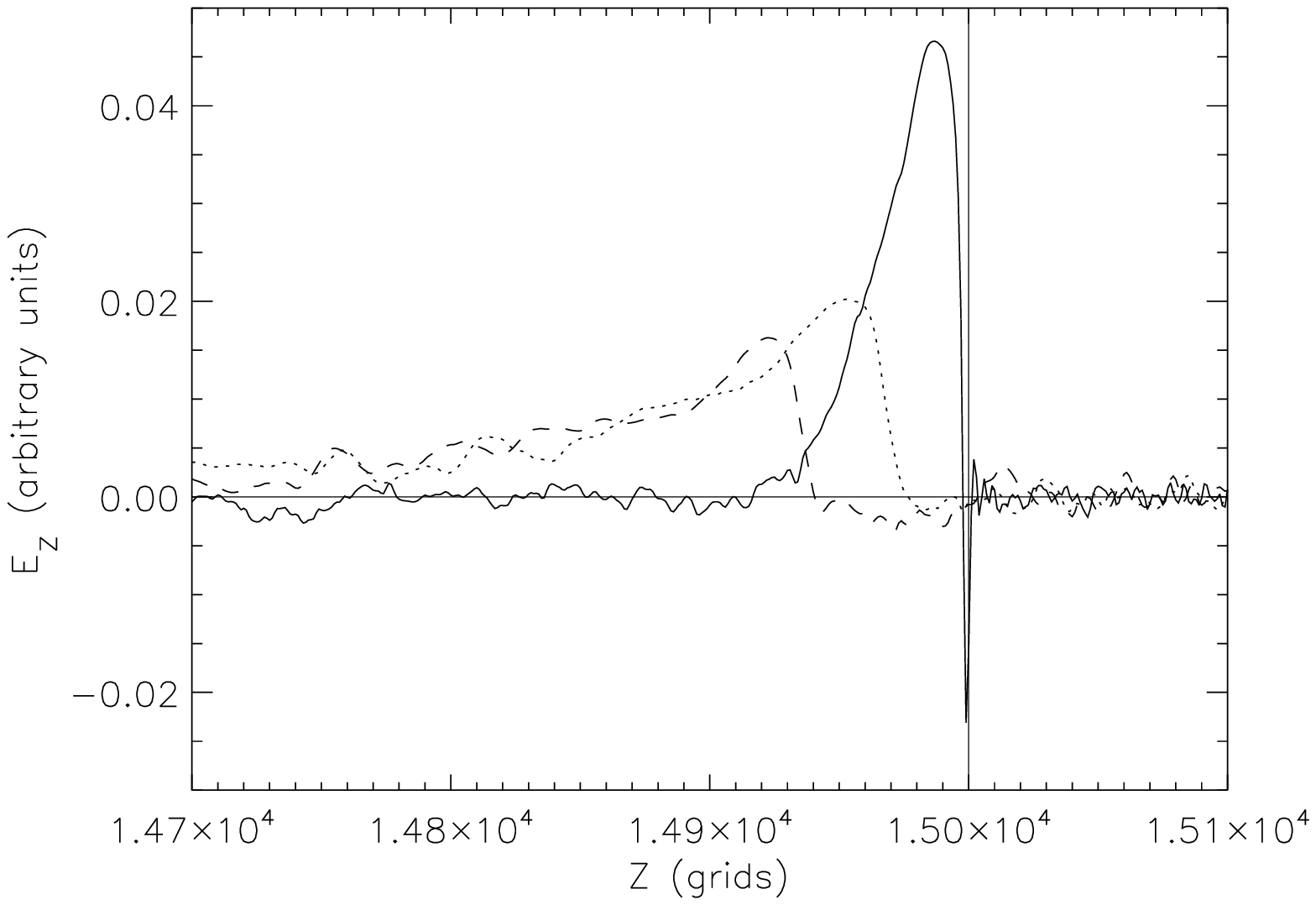}
\includegraphics[width=8cm]{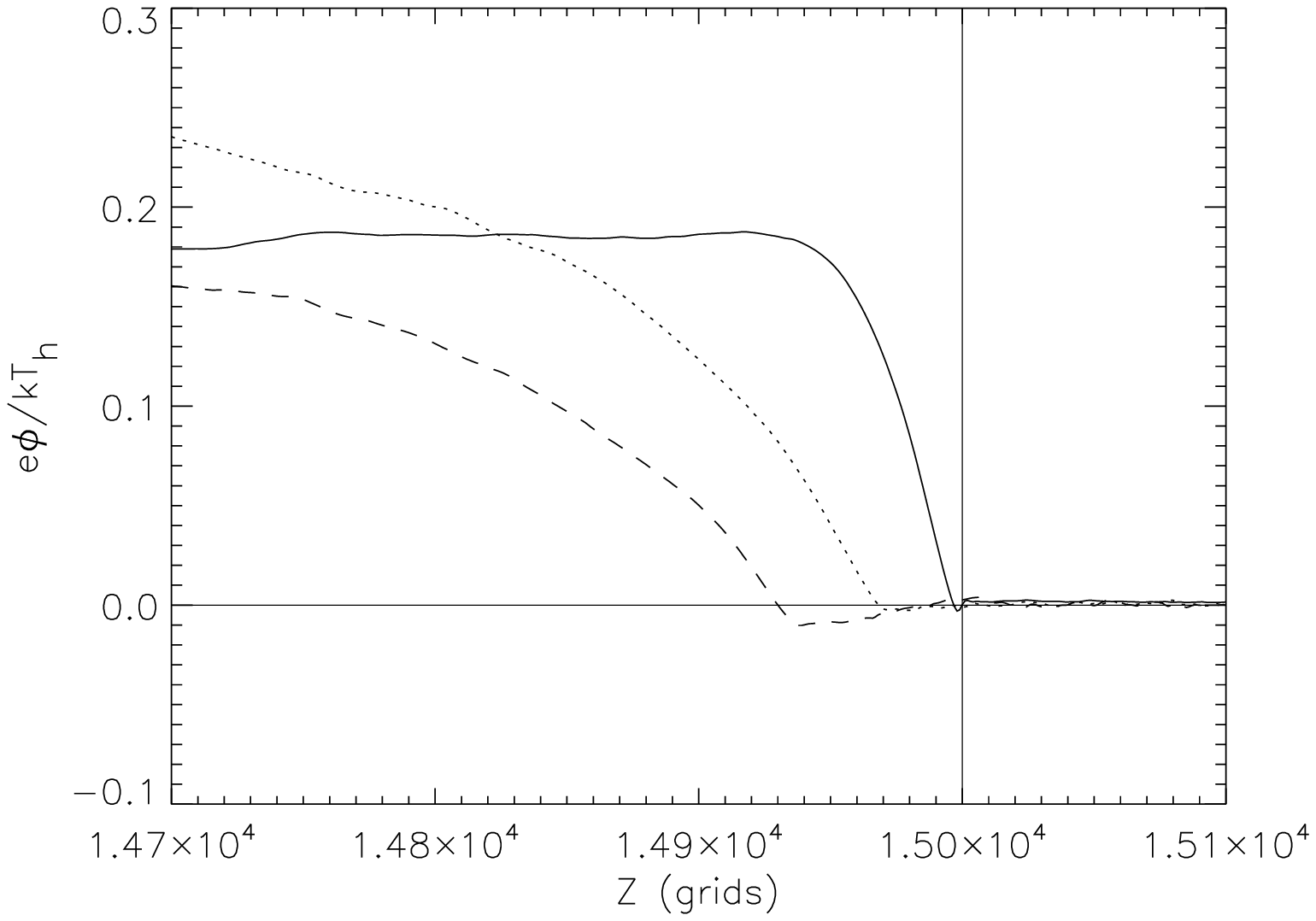}
\end{center}
  \caption{Upper part: Electric field in the z-direction
  in the space interval 14700 - 15100 grids at $\omega_{pe}t$ = 25 (solid line),
   at $\omega_{pe}t$ = 500 (dotted line), and at $\omega_{pe}t$ = 1000 (dashed line).
   Bottom part: Electric potential in the z-direction in the same space interval and same times.
   (Run with electromagnetic interactions.)}
  \label{figure9}
\end{figure}

\section{NUMERICAL MODEL}

In the model we study an interaction between the hot (rare) and cold (dense)
plasma. We consider the collisionless case, i.e., the case with the
particle-wave interactions only.

We use a three-dimensional (3-D) electromagnetic relativistic particle-in-cell
(PIC) model (Karlick\'y 2009) in order to include the both electrostatic and
electromagnetic effects. Namely, in our study of the similar problem (problem
of the thermal conduction front) we recognized an importance of the
electromagnetic processes \citep{2015ApJ...814..153K}.

The system size is $L_x$ = 8$\Delta$, $L_y$ = 8$\Delta$ and $L_z$ =
30000$\Delta$ (where $\Delta$ is the grid size). Thus, the system is
effectively one-dimensional along the assumed magnetic field. In the model we
initiate the electron-proton plasma with the proton-electron mass ratio
$m_p/m_e$=100. This value is chosen to accelerate processes connected with
protons. Nevertheless the ratio is still sufficient to clearly separate the
dynamics of electrons and protons.

The numerical system is divided into two parts with the hot (rare) plasma
(located at $z$ = 0 -- 15000$\Delta$) and cold (dense) plasma ($z$ = 15000 --
30000$\Delta$). In the initial state the velocities of particles in the both
parts of the system are described by the Maxwellian distribution functions. In
the colder plasma the electron thermal velocity is chosen as $v_{Tec}$ = 0.025
$c$, where $c$ is the speed of light. The corresponding temperature is
\textit{T}$_{\rm c}$ = $v_{Tec}^2 m_e/(2 k_B)$ = 1.85 MK, where $k_B$ is
Boltzmann constant.  On the other hand, the the electron thermal velocity in
the hot plasma is 10 times greater (the temperature is 100 times greater =
T$_h$ = 185 MK) then that in the cold plasma. Thus, the ratio of temperatures
in the hot and cold plasmas agrees to that in the solar transition region, but
the temperatures are much higher then in the corona and chromosphere. Such a
selection of temperatures is given by requirements of the PIC modelling, namely
the electrostatic and electromagnetic waves in the system need to be correctly
described. The initial temperatures of protons in the both parts of the system
are the same as for electrons.

To simulate conditions as in the solar transition region with the pressure
equilibrium we take 30 electrons and 30 protons per cube grid in the hot plasma
part and 3000 electrons and 3000 protons per cube grid in the cold plasma part
of the system. Note that this difference in particle densities and extension of
the system for sufficient distances for particle propagations require very
powerful computer.

The plasma frequency in the cold and dense plasma is $\omega_{pe}$ = $2\pi/t_p$
= 0.05, where $t_p$ is the plasma period. The magnetic field is oriented in
$z$-direction and its value corresponds to the electron gyro-frequency
$\omega_{ce}$ = 0.1 $\omega_{pe}$. In the hot (rare) plasma the plasma
frequency is ten times lower than that in the cold (dense) plasma. The time
step is $\Delta$t = 1. The electron Debye length in the hot plasma part is
$\lambda_D$ = 2.5 $\Delta$ and in the cold plasma part $\lambda_D$ = 0.25
$\Delta$, respectively. In the $x$- and $y$-directions the periodic boundary
conditions are used. In the $z$-direction we used free boundary conditions.

Computations were performed at the IT4Innovations National Supercomputing Center, Ostrava, Czech Republic.

\section{Results}

To understand processes in the transition region between hot and cold plasma we
made two types of computations: a) without any interactions (free expansion of
particles) and b) with the electrostatic and electromagnetic (particle-wave)
interactions.

Results of the computation with the free expansion of particles are presented
in Figures~\ref{figure1} - \ref{figure4}. Figure~\ref{figure1} shows a time
evolution of the electron density, normalized to the initial density in the hot
plasma, along the $z$-coordinate in the region close to the temperature
(density) jump, at times $\omega_{pe}t$ = 500, $\omega_{pe}t$ = 1500, and
$\omega_{pe}t$ = 2500, where $\omega_{pe}$ here and in the following means the
plasma frequency in the cold (dense) plasma. The upper and bottom parts of the
figure show the evolution of the electron and proton densities. The
corresponding evolution of the electron energies (pseudo-temperatures because
particle distributions near the transition region deviate from the Maxwellian
ones) is shown in Figure~\ref{figure2}. Although the thermal velocities of
electrons and protons in the hot plasma are in the initial state 10 times
greater than those in the cold plasma, electrons as well as protons mainly flow
from the colder (denser) plasma to hotter (rarer) plasma. It is due to that the
electrons and protons in the cold plasma are 100 times more numerous, see the
negative electron and proton flux in Figure~\ref{figure3}. Thus, the electron
and proton densities at the hotter part of the system close to the transition
region increase and their temperatures decrease. Because the velocities of the
cold plasma electrons are greater than those of the cold plasma protons, the
electron flux is greater the the proton flux (Figure~\ref{figure3}). See also
that the cold plasma electrons penetrate deeper into the hot part of the
system. It is due to that in such computations there are no particle-particle
and particle-wave interactions.

But, simultaneously with the above described process the fast electrons from
the hot part of the system penetrate into the region with the cold plasma, see
the time evolution of the tail of the distribution function of electron
velocities in the $z$ direction in space $z$ = 15000 -- 16000 grids
(distribution tails in the velocity interval $v_{ez}/c$ = 0.1 -- 0.3)
(Figure~\ref{figure4}, upper part). On the other hand, in Figure~\ref{figure4}
(bottom part) we can see a penetration of cold plasma electrons into the hot
part of the system; compare the distribution functions in the initial state and
those in the following times. The decrease of the hot plasma electrons with the
negative velocities ($v_{ez}/c$ = -0.3 -- -0.1) in the region $z$ = 14000 --
15000 grids, at $\omega_{pe}t$ = 500 (dotted line) and at $\omega_{pe}t$ = 2500
(dashed line) (Figure~\ref{figure4}, bottom part) is caused by an escape of the
electrons with such velocities from this region to the region with more
negative locations and by a shortage of the electrons with such velocities on
the cold side of the transition region.

Now, let us compare these results with those computed with the electrostatic
and electromagnetic interactions (Figures~\ref{figure5} - \ref{figure8}).
Similarly as in the case without any interactions there is the negative
electron and proton flow from the cold to hot part of the system
(Figure~\ref{figure7}). However now, contrary to the previous case, the
electron and proton flux is nearly the same and corresponds roughly to the
proton flux as in the case with the free expansion of particles. It is due to
that the plasma tries to keep the constant electric current density (in our
case the zero initial current density). In the initial state at the transition
region the flux of the cold plasma electrons towards the hot plasma is higher
than that of the proton flux. Thus, in the very short time at the head of the
proton flow a region with a surplus of the negative charge (cold electrons are
faster than cold protons) appears and also here the non-zero electric current
is formed. Both these electrostatic and electromagnetic effects cause that the
protons at this region are accelerated and electrons decelerated in the
$z$-direction, in such a way that the both electron and proton fluxes are kept
nearly the same (i.e., with zero electric current). Note that in the
beam-plasma system the same process leads to a formation of the so called
return-current \citep{2009ApJ...690..189K}. For details of these processes, see
\cite{1990A&A...234..496V}.

Because protons are much more heavier than electrons the expansion of the cold
plasma electrons into the hot part of the system is much slower than in the
case without any interactions, compare Figure~\ref{figure5} and
Figure~\ref{figure1}. Thus also a decrease of the electron temperature in the
hot part of the system close to the transition region is slower than in the
previous case, see Figure~\ref{figure6} and Figure~\ref{figure2}.

Furthermore, Figure~\ref{figure8} shows time evolution of the electron
distribution function in the $v_{ez}$ velocities at the both sides of the
hot-cold transition region. As seen here and in comparison with
Figure~\ref{figure4}, the electromagnetic processes, described above, reduce
the penetration of cold plasma electrons into the hot part of the system as
well as the penetration of hot plasma electrons to the cold plasma region;
compare Figures~\ref{figure8} and \ref{figure4} (upper parts) in the velocity
interval $v_{ez}$ = 0.1 -- 0.3. Even some hot plasma electrons are
backscattered, see the distribution expressed by the dashed line for the
velocities around $v_{ez}/c$ = - 0.08. This effect indicates that around the
hot-cold transition region the plasma waves are generated which backscatter the
hot plasma electrons. Another and even stronger effect of the backscattering of
electrons can be seen by comparing of Figures~\ref{figure8} and \ref{figure4},
bottom parts. While in the case without any interactions the cold plasma
electrons penetrating to the hot part of the system have only negative
velocities, in the case with the electromagnetic interactions the cold plasma
electrons have also positive velocities. The maximum of the distribution of
these cold plasma electrons is slightly shifted to the negative velocities.
This velocity shift corresponds to the velocity of the distribution maximum of
the cold plasma protons flowing together with the electrons from cold plasma to
hot plasma. The cold plasma protons in the hotter part of the system have only
negative velocities and for this short time of evolution they are only
negligibly scattered.

Figure~\ref{figure8} (bottom part) also shows that the electron distribution in
space around the hot-cold transition region is a mixture of the hot and cold
plasma electrons.

We also analyzed the electric field and corresponding potential in this
hot-cold plasma system. We found that except the negatively oriented electric
field at the very beginning of the system evolution and in the space interval
14998-15000 grids (Figure~\ref{figure9}, solid line in the upper part), the
electric field is positively oriented and decreases in time
(Figure~\ref{figure9}, the upper part). In the bottom part of
Figure~\ref{figure9} we present the corresponding normalized electric
potential, which is about $e \phi/ k_B T_h$ = 0.2, where $e$ is the electric
charge, $\phi$ is the potential, $k_B$ is the Boltzmann constant, and $T_h$ is
the temperature of the hot plasma.

\section{Discussion and conclusions}

Using the 3-D PIC model we studied processes at the transition region between
hot (rare) and cold (dense) plasma. Ratios of the initial hot and cold plasma
temperatures and rare and dense plasma densities were taken as 100 and 0.01,
respectively, i.e., as in the case with the pressure equilibrium in the solar
atmosphere. Two types of computations were made: with and without the
electromagnetic interactions.

In both the cases we found that the flow of cold plasma electrons and protons
from colder plasma to hotter one is greater than that of hot plasma electrons
and protons in the opposite direction. It is given by the initial conditions in
the system, where the relation $n_c T_c = n_h T_h$, where $n_c$ and $n_h$ are
the particle densities and $T_c$ and $T_h$ are the cold and hot plasma
temperatures, is used. Note that the $T_c$ and $T_h$ are proportional to the
square of the thermal particle velocities $v_{Tc}$ and $v_{Th}$, and the
particle flux is $n_c v_{Tc}$ and $n_h v_{Th}$, respectively.

Due to this flow the plasma in the hotter part of the system becomes colder and
denser during time evolution. In the case without any interactions among
particles cold plasma electrons and protons freely penetrate into the hot
plasma. However, the cold plasma electrons are faster than cold plasma protons
and therefore they penetrate deeper into the hotter part of the system than the
protons.  Therefore, the cooling of the electron and proton components of the
plasma in the hotter part of the system is different. On the other hand, in the
case with the electrostatic and electromagnetic interactions, owing to the
plasma property which tries to keep the total electric current constant
everywhere (in our system close to zero), the cold plasma electrons penetrate
into the hotter part of the system together with the cold plasma protons. Thus,
the cooling of the hotter plasma is much slower and it is the same for
electrons and protons. Note that for the real proton-electron mass ratio (1836)
the process of penetration of the cold plasma protons and electrons into the
hotter part of the system will be even slower than in the present case, where
we used the mass ratio equal to 100. Furthermore, the plasma waves generated
during these processes reduce a number of the electrons escaping from the hot
plasma into the colder one and scatter the cold plasma electrons penetrating
into the hotter part of the system.

We found that except the negatively oriented electric field at the very
beginning of the system evolution and in the very narrow region close to the
boundary between the hot and cold plasma, the electric field is oriented in the
opposite direction comparing to that of the expanding cold plasma to hotter
one, and the electric field decreases in time.

Solving Vlasov equations
~\cite{1982JGR....87.9154S,1987LPB.....5..233S,2011PhPl...18l2105S} studied an
expansion of the plasma into vacuum and expansion of the plasma into the plasma
with lower density, but having the same temperature. In the both these cases
the electric field was oriented in the direction of plasma expansion. It is due
to that the electrons expand faster than protons. In our simulations we
recognized this effect only at the very beginning of the system evolution and
only in the very narrow region close to the boundary between the hot and cold
plasma. But simultaneously, the another effect caused by hot plasma electrons
appears. Namely, the hot electrons in the hot plasma region close to the
hot-cold boundary expand to the colder plasma side and thus at this region a
surplus of the positive charge appears. It produces the found electric field
oriented in the opposite direction comparing to that of the expanding cold
plasma to hotter one.

If we compare the processes in the present model with those in the thermal
front, computed in the plasma having the constant density,
\citep{2015ApJ...814..153K}, we can see that only the process reducing a
penetration of hot plasma electrons into the cold plasma (due to the plasma
waves) is similar. The main difference is given by a strong jump in densities
of the hot and cold plasmas and corresponding flow of the cold plasma protons
(together with the cold plasma electrons) from the cold (dense) plasma to
hotter (rare) one.

Now a question arises if this cold proton (cold electron) flow can be stopped
by some way and thus stabilize the temperature-density structure as in the real
transition region in the solar atmosphere. Analyzing this problem, we think
that this flow can be stopped by the gravity force together with the generated
plasma waves. Namely, the flow of the cold plasma electrons and protons in the
real solar atmosphere is oriented upwards, thus the gravity force can
decelerate and stopped this flow.

Furthermore, we think that the temperature jump in the transition region
(between the corona and chromosphere) needs some more or less permanent source
of heating in the corona (probably nanoflares). On the other hand, the flow of
hot plasma electrons from the corona to the chromosphere is reduced by the
plasma waves generated at the transition region, as shown in these simulations.
Based on our analysis of thermal fronts~\citep{2015ApJ...814..153K} we think
that the flow of hot electrons to the transition region can be even more
reduced, if the strong ion-sound turbulence is present at the transition
region.

Finally, we found that the electron distribution in space around the hot-cold
transition region is a mixture of the hot and cold plasma electrons. This can
be important in interpretations of the X-ray line spectra formed at the solar
transition region \citep{2017A&A...603A..14D,2017arXiv170603396D}.

\begin{acknowledgements}
We acknowledge support from Grants 16-13277S and 17-16447S of the Grant Agency
of the Czech Republic and institutional support from the University of Ostrava (IRP201557).
This work was supported by The Ministry of Education, Youth and Sports from the Large Infrastructures for Research, Experimental Development and Innovations project „IT4Innovations National Supercomputing Center – LM2015070“.
\end{acknowledgements}

%

%


\end{document}